\documentclass{iopconfser}
\usepackage{amsmath}
\begin{document}

\title{Physical complexity and black hole quantum computers}

\author{Michele Reilly, Seth Lloyd}

\affil{Department of Mechanical Engineering, MIT, Cambridge MA, USA}

\email{mmreilly@mit.edu, slloyd@mit.edu}







\begin{abstract}


The ultimate limits of computation are not just logical, but physical. We investigate the physical resources -- time, energy, entropy, and free energy -- required to perform computational work. We apply the resulting measures of physical complexity to conventional electronic computers, to quantum computers, to biological systems, to black holes, and to the universe itself, with implications for artificial intelligence development where biological efficiency limits suggest new computational paradigms beyond current digital architectures.

\end{abstract}

\section{Introduction}

\bigskip

The theory of computational complexity is based on the trade-off between two resources, computer time and memory space.   Computer time is defined to be the number of elementary logical operations required to perform a computation.  Memory space is the number of bits of information required by a computer over the course of a computation.   
\\
  
This paper investigates the physical counterparts of temporal and spatial complexity.  
We define the physical analogue of time computational complexity to to be proportional to the time it takes a physical system to perform a computation.    We use the fundamental physics of computation [1-2] to relate physical time to the number of elementary logical operations. 
Logical operations flip bits, and flipping bits requires both time and energy.    The more energy one puts into flipping a bit, the faster it can flip. We identify physical analogue of computational time with the number of bit flips that can take place over time $t$ given energy $E$: 
$\Phi$-TIME $ = 2Et/\pi\hbar$. The numerical factors are given by the Margolus-Levitin theorem [1-2], which limits how fast a quantum system flip a bit given energy $E$.  
\\

Now turn to the physical analogue of memory space -- the number of bits of information that a computer requires access to in the course of a computation.
Physically, information is intimately connected with entropy [1-2].  Entropy is the number of bits of information required to specify the microscopic state of a physical system. 
At thermal equilibrium, i.e., maximum entropy, these bits are random.
Computation requires access to non-random bits, however. The theory of reversible computation [1-9] suggests that we should define the physical analogue of the amount of memory space required for a computation,
$\Phi$-SPACE,
to be the computing system's negentropy, i.e., the difference between the maximum entropy $S_{max}$ of the system
and its actual entropy, $S$: $S_{max} - S$ is an upper bound to the amount of memory space required by the computation, either reversible or irreversible. 
$\Phi$-SPACE is the maximum number of bits of memory space available for a physical system to perform computation.
\\

Computations take not only time and space: they also take work.  
The amount of work $W$ that can be performed by a physical system is equal to $F-F_{eq}$, where $F=E-TS$ is the free energy and $F_{eq}$ is the system's free energy when brought to thermal equilibrium at temperature $T$.  
Every logical operation requires work/free energy to perform, and some fraction of that free energy is inevitably dissipated in each logical operation.   As a result, the free energy used doing a computation is proportional to the number of operations performed.  Moreover, as we will show below, the free energy also determines the amount of physical memory space available.   
Because of the central role free energy plays as the physical resource that determines how much work can be done, how many ops can be performed, and how much memory space is available we
define the {\it free physical complexity} FREE-${\Phi}$ to be the amount of free energy used during the computation. 
\\

The following table summarizes the relationship between conventional computational complexity and physical computational complexity:

\begin{center}
\begin{tabular}{|l|l|}
\hline
\textbf{Computational Complexity} & \textbf{Physical Complexity} \\
\hline
TIME = \# ops & $\Phi$-TIME = ${2 E t}/{\pi \hbar}$ = accumulated quantum phase \\
SPACE = \# bits & $\Phi$-SPACE = $S_{\text{max}} - S$ = negentropy/thermodynamic depth \\
\hline


\end{tabular}

\vspace{1em}

\end{center}
 
Physical spatial and temporal complexity provide upper bounds on the amount of computation that can be performed by any physical system.    
We discuss the implications of physical complexity
for the computational complexity hierarchy. In particular, we show
that in the presence of noise and errors in the computation,
$\Phi$-SPACE is proportional $\Phi$-TIME, and both are proportional to FREE-$\Phi$. Accordingly, a natural definition of physical computational complexity with the amount of free energy required to perform a computation.  We apply the resulting theory of physical computational complexity to
analyze the computational power of black holes: exactly half of the
energy of a black hole is free energy that is available for physical computation.


\section{Background}  

\bigskip
In 1936, Alan Turing invented
the Turing Machine, a machine
which defined the concept of universal computation [11].
Crucial to Turing's conception was the fact that the universal Turing
machine was consistent with physical reality: at any point in the computation,
the number of logical steps that has been performed by the machine is finite, as is the
amount of memory space employed.   An important subtlety is that Turing
allowed the always finite computation to be indefinitely extensible:
we can imagine that the length of the computation and the amount of
memory space used could increase without limit.
The development
of actual, physical computers during the 1940s and 1950s raised the
question of what were the physical limits to the resources
required by computation.
\\

In 1949 John von Neumann suggested that each
logical operation required a minimum of $k_B T$ worth of energy to
be dissipated to the environment at temperature $T$ [12].
In 1962, Rolf Landauer showed
that only {\it logically irreversible} operations required dissipation [3].
The microscopic reversibility of the laws of physics implies that
an irreversible logical operation such as AND, OR, or ERASE, that irreversibly erases
a bit from the system must transfer that bit to the environment,
resulting in entropy increase of $k_B \ln 2$ and energy dissipation
$k_B T \ln 2$.  Landauer's principle was followed by the work of Lecerf
[4] who showed that Turing machines could be made logically reversible.
Lecerf's paper attracted little notice (according to Google Scholar it was cited exactly once between 1963
and 1992), and in 1973, Charles Bennett independently
developed a full theory of reversible
Turing machines based on the DNA repair mechanism [5-6].   In 1982,
Ed Fredkin and Tom Toffoli presented a ballistic model of computation
based on reversible logic gates [7].   At the same time, Paul Benioff presented
a reversible quantum model of computation [8].
\\

The set of
computations that Benioff's quantum model could perform were simply those that could be performed
by a classical Turing machine.   In 1985, David Deutsch showed how one
could implement a universal quantum Turing machine by supplementing the
classical logic gates with operations that put quantum bits in superposition [9].
Reversible classical computation was a necessary precursor to quantum computation: only if the classical logical operations are performed reversibly can the quantum coherence in the superpositions that make up a quantum computation be preserved.

\bigskip\noindent{\it Additional memory requirements for reversible computation}

\bigskip
Embedding an irreversible logic operation such as AND or OR in a reversible logic gate such as a Fredkin or Toffoli gate requires extra bits of input and output.  Naively, then, it might appear that reversible computation requires extra bits of memory space for each logical step of the computation.   Bennett showed, however, that the additional bits can be reused and recycled in a clever way that expands the additional memory space required for reversible computation by only a constant multiplicative factor over the memory space required for irreversible computation [5-6].    Consequently, at the abstract level, the computational complexity classes P and PSPACE (representing the classes of computations that can be performed in polynomial time or in polynomial space as a function of program length) remain unaltered in the reversible context.  
\\

At the physical level, however, the relationship between temporal and spatial computational complexity is different.  In particular, any physical computational device will inevitably suffer from noise and errors.   Error-correcting codes can be used to identify and correct errors, allowing one to perform arbitrarily long computations even in the presence of noise.   In the context of reversible and quantum computation, error correction requires fresh, zero-entropy ancilla bits/qubits to be supplied to the computer at each stage of the error correcting process, and rejects noisy, random bits to the environment, erasing them from the computation.
\\

Each bit of noise rejected to the environment increases the entropy of the environment by one bit, and requires free energy $k_B T ln 2$ to be dissipated, by Landauer's principle.   Alternatively, the noisy, entropic bits specifying which errors have already occurred can be collected in the reversible computer's memory.
In either case, no matter how small the error rate per logical operation, the amount of physical, low entropy bits required to perform the computation -- the physical spatial complexity or logical depth -- grows linearly with the length of the computation.

\section{Physics of computation}

\bigskip
We now give a summary of the basic results of the fundamental physics of computation [1-2].   We then apply these results to analyze the physical analogues of spatial and temporal computational complexity.

\subsection{Physical memory space} 

\bigskip

Computation requires memory space, which in turn requires degrees of freedom that are not in their maximum entropy state.  If we have a quantum system with $n$ degrees of freedom, and total entropy $S \leq S_{max}$, where $S_{max}$ is the system's
maximum entropy for a degree of freedom, then in general as $n$ becomes large we can reversibly transform the mixed state of the system to obtain a subset of the degrees of freedom in a pure state, containing $ \approx S_{max} - S $ bits/qubits, with the remainder of the system in an almost fully mixed state containing $\approx S$ bits/qubits [1].
\\

That is, the total number of bits of memory available to a system with entropy $S$ is no greater than $S - S_{max}$, and we can generically reversibly obtain 
$ \approx S_{max} - S$ pure bits/qubits of information from that state.
The quantity $S_{max} - S$ is called the negentropy of the system: negentropy measures how far the system is away from its maximum entropy.   Motivated by the observation that the number of zero-entropy bits available for a reversible computation, classical or quantum, is bounded by the negentropy, Lloyd and Pagels [10] defined the thermodynamic depth of a physical process such as a computation to be equal to the negentropy of the  process.  Accordingly, we identify the physical spatial computational complexity of a computation, $\Phi$-SPACE, with the minimum negentropy/thermodynamic depth of a physical system that performs the computation.
\\

Computations performed by actual physical computers are highly dissipative and use much more negentropy than the minimum required by the laws of physics.   A 5nm transistor has a capacitance of approximately $0.1$ femtofarad, and dissipates approximately $10^{-16}$ joules of free energy per irreversible logical operation.   By contrast, $k_B T$ at room temperature is approximately $5 \times 10^{-21}$ joules.   The negentropy of a physical bit in conventional memory, is similarly around five orders of magnitude higher than $k_B \ln2 $.   
\\

Looking beyond human-built digital computers to information processing in living systems in general, we can identify a `bio-op' with a the formation or breaking of a molecular bond in the chemical reactions involved in the information intensive processes of DNA replication, reproduction, and metabolism.  The amount of free energy associated with breaking/forming covalent bonds in biomolecules is on the order of one electron volt $=1.6 \times 10^{-19}$ joules.    Hydrogen bonds have energy an order of magnitude lower, and van der Waals bonds have energy lower still.
An individual human being expends $\approx 100$ watts of energy ($\approx 100$ kilocalories/hour) on reproduction and metabolism Accordingly, cellular information processing in an individual human being is performing on the order of $10^{20-22}$ bio-ops per second  --  approximately the same as the number of logical operations performed by all the electronic computers in the world (one zetaflop).   

\subsection{\it Error correction}

\bigskip

To measure the amount of physical resources required for computation, we need to include the negentropy/free energy required for error correction.   
A bit-flip error that occurs with probability $\epsilon$, injects  entropy $ \Delta S = -\epsilon \log \epsilon - (1-\epsilon) \log (1-\epsilon) \approx \epsilon \log (1/\epsilon) $ into the bit.  Error correction 
protocols restore the bit to its proper value while pumping that entropy into the environment at a free energy cost of $k_B T \Delta S$ (Landauer's principle).
\\

Phrased in terms of reversible and quantum computing, the physical spatial complexity measures not only the information/negentropy of the computer itself, but the information/negentropy of the fresh ancilla bits that must be added to the computer to perform the error correction. 
Consequently the physical spatial computational complexity required to perform a computation with $t$ steps and error rate $\epsilon << 1$ per step goes as $t \epsilon \log (1/\epsilon)$: physical spatial computational complexity grows at least proportionally to the error rate times number of logical steps in the computation.
\\

By contrast, conventional spatial computational complexity corresponds only to the maximum negentropy of the computer's memory during the computation, and not to the full thermodynamic resources used by the computational system, which includes the supply of fresh ancilla bits and the erasure of used ancillae.

\subsection{Physical computation time}

\bigskip

Digital computation, both classical and quantum, is described by a sequence of logical operations (logic gates) applied to the degrees of freedom of the system performing the computation. The number of logic gates in a computation is the computation's time computational complexity.
\\

Physically, each logical operation corresponds to a dynamical process.  In a conventional electronic computer, logic gates are constructed of transistors that dynamically change their output state as a function of the states of their inputs.  In a quantum computer, a logic gate corresponds to a unitary transformation acting on a few degrees of freedom.
Similarly, in biological information processing, as noted above, we can define an operation as the formation or breaking of chemical bonds (covalent, hydrogen, van der Waals) in the service of reproduction and metabolism.
\\

The ultimate physical limits to computation are described in [1-2].  A logic gate must be able to change the logical degrees of freedom of the computer from one state to another, e.g., flipping a qubit from the logical state $|0\rangle$ to the logical state $|1\rangle$, and {\it vice versa}.  Formally, $|0\rangle$ and $|1\rangle$ are vectors in a two-dimensional complex space, and the quantum bit flip is accomplished by applying a Hamiltonian $H = -E\sigma_x$ for an amount of time $\Delta t=\pi \hbar/2E$.
\\

During the bit flip process, the average energy of the qubit above its ground state is $E$,
which is also equal to its spread in energy $\Delta E$.
The amount of energy required to flip the bit in time $\Delta t$ thus obeys $E = \pi \hbar/2 \Delta t$, which is also equal to the spread in energy required to flip the bit in that time. 
These relations also apply to quantum mechanical implementations of Toffoli and Fredkin gates, reversible gates which suffice for universal reversible computing (and indeed to any reversible logical operation which is its own inverse).
\\

More generally, it can be shown that {\it any} quantum mechanical process that performs a reversible logic operation capable of flipping a bit in time $\Delta t$ obeys $E\geq \pi\hbar/2\Delta t $, where $E$ is the expectation value of the system's energy minus its ground state energy (the Margolus-Levitin theorem/quantum speed limit [1].   Similarly, $\Delta E \geq \pi\hbar/2\Delta t $ (the Heisenberg limit).
\\

As a consequence of the Margolus-Levitin theorem,   the total number of logical operations that can be performed by a physical system with energy $E$ above its ground state obeys
$ \# ops \leq 2Et/\pi \hbar.$
Moreover, quantum computers saturate this bound if $E$ is taken to be the energy of the time-dependent interactions used to apply the quantum logic gates in the computation.
\\

Because each logic gate involves moving millions of electrons, conventional electronic computers operate much less efficiently than the bound of the theorem above.   Similarly, biological information processing operates much more slowly than the bound. An exception is photosynthesis, where photon absorption and exciton diffusion are intrinsically quantum-mechanical processes that can saturate the Margolus-Levitin bound.
\\

Recognizing the diversity of different types of information processing, and noting that quantum computers can come close to the physical limits of the theorem, we identify the physical temporal complexity $\Phi$-TIME of a computational process -- such as solving a well-defined problem -- with the minimum energy-time product
$2Et/\pi\hbar$ over the set of physical processes that solve that problem.   Since quantum computers saturate the physical limit for performing logical operations, the minimum energy-time product for a computation is proportional to the conventional temporal computational complexity.
\\

Note that, like conventional spatial and temporal computational complexity, $\Phi$-SPACE and $\Phi$-TIME are just numbers -- but numbers that are dictated by the laws of physics {\it together} with the laws of computation, rather than by the laws of computation alone.  Unlike conventional spatial and temporal computational complexity, physical spatial and temporal computational complexity have a direct physical interpretation in terms of the resources of negentropy, energy, and time required to compute.

\section{ Free physical complexity}

\bigskip
The two physical measures of complexity proposed so far are the physical analogues of spatial computational complexity (logical depth) and temporal computational complexity (energy time product).   The necessity to correct errors -- no matter how small the rate that they occur -- implies that $\Phi$-SPACE grows at least proportionally to $\Phi$-TIME, for any class of physical computing systems.
Before exploring further the tradeoffs between physical spatial and temporal complexity, we introduce a new measure, free physical complexity, which combines both spatial and temporal complexity in one measure.
\\

Recall that the free energy of a system with energy $E$, entropy $S$, and temperature $T$, is
$F = E - TS$.    The equilibrium free energy for a system with Hamiltonian H is obtained by maximizing the entropy $S = - {\rm tr} \rho \log \rho$ for a fixed expectation value of the energy, 
$E = {\rm tr} H \rho$.   The thermal equilibrium state is a Gibbs state, $\rho_{th} = Z^{-1} e^{-\beta H/k_bT}$, where $T$ is chosen to satisfy ${\rm tr} \rho_{th} H = E$.   The partition function $Z={\rm tr}~ e^{-H/k_bT}$ normalizes $\rho_{th}$ to 1.
\\

The free energy can be increased by increasing the system's energy, or by decreasing its entropy, or both.    Free energy is essentially energy that is unencumbered by entropy, and that is thus available to do work. In particular, the non-equilibrium free energy
for a quantum state $\rho$ is defined to be 
$F(\rho) = E - TS$, where $E = {\rm tr} \rho H$, and $S = - {\rm tr} \rho \ln \rho$.   Non-equilibrium free energy governs the amount of work that can be extracted from a system by returning it to its equilibrium state, while putting it into contact with a thermal bath at temperature $T$.    The amount of work that can be extracted from the system
is equal to 
$W(\rho) = F - F_{eq} = T ( -{\rm tr} \rho \log \rho_{th} + {\rm tr} \rho \log \rho).$
\\

That is, free energy governs the amount of work that can be extracted from a system that contains both ordered energy, and disordered energy in the form of heat.  Work is energy that is free from entropy.   Quite generally, we can use reversible, energy-preserving dynamics to transform a multi-variable system so that it divides into two sets of variables or parts.   The first part contains all the entropy and has zero extractable work.  The second part has zero entropy, and contains all the extractable work.   Indeed, we can extract that work by applying a unitary transformation that transforms the pure state of the second part into the ground state, extracting its energy as work in the process. 
\\

We define the free physical complexity,
FREE-$\Phi$, of a system performing computation to be simply the free energy used by the system over the course of the computation. 
Free physical complexity (``usable energy-based complexity" or ``work-based complexity") combines spatial and temporal physical complexity in a natural and physically motivated way.  Note that  Free-$\Phi = T(S_{eq} - S(\rho))/\pi\hbar$, where $S_{eq}$
is the maximum entropy for a state with energy $E$ and $T$ is the temperature of the corresponding thermal state.
That is, FREE-$\Phi$ is proportional to the system's effective temperature times the thermodynamic depth of the system [10].
\\

Free physical complexity combines both memory space and the number of ops into a unified physical complexity measure with a definite physical interpretation: free energy is the ``currency of intelligence” in the universe.   Most important, free physical complexity is directly applicable to characterizing the physical computational complexity of physical processes that go beyond conventional computation.  In biological information processing, for example, free physical complexity is simply proportional to the total free energy put to use in the forming and breaking of chemical bonds that makes up the `computation' that is the life of an organism.
\\

While we have phrased our argument so far in terms of quantum computers, physical spatial and temporal complexity, together with free physical complexity, apply to any well-defined finite physical process.   
Consider, for example, a sequence of chemical reactions that produce a complex organic molecule.    We start from a few types of simple molecules, which undergo a reaction to produce more complex (in the sense of higher molecular weight) molecules.   This first reaction may happen spontaneously, driven by the consumption of free energy present in the molecules themselves, or may be facilitated by the injection of additional free energy, for example, in the form of photons from the sun.   Then the more complex molecules interact with each other, together with simpler molecules and additional injections of free energy to produce more complex molecules (again as measured by molecular weight).  After multiple steps, the final, complex chemical compound is produced.
\\

In such chemical processes, a good measure of complexity is the assembly index [13], equal to the minimal number of distinct chemical reaction steps required for the assembly of that chemical compound.   Free physical complexity supplements the assembly index by including 
the amounts of free energy required to perform each of those steps.  Free physical complexity participates in the advantages of highly useful assembly theory descriptions of chemical processes, while adding extra information in the form of the amounts of free energy and time required to produce each intermediate compound.
\\

\section{Implications for Artificial Intelligence Development}

\bigskip
The physical limits of computation described above have immediate implications for the development of artificial intelligence systems. Current AI architectures operate far from thermodynamic efficiency: training large language models requires megawatts of power while biological neural networks achieving comparable performance operate on $\sim$20 watts. This $10^5$ efficiency gap suggests fundamental architectural limitations in digital approaches rather than proximity to physical limits.
\\
\\
Three critical trade-offs emerge from our analysis. First, the energy-time relationship $\Phi$-TIME $= 2Et/\pi\hbar$ reveals an inescapable trade-off between computation speed and energy efficiency. Current AI systems optimize for speed at enormous energetic cost, but biological systems demonstrate that slower, more energy-efficient computation can achieve superior performance through parallelism and fault tolerance.
\\
\\
Second, the necessity of error correction creates a fundamental tension between memory requirements and error tolerance. As AI systems scale toward longer-running autonomous agents and continuous learning, error accumulation becomes the dominant constraint. The physical spatial complexity grows linearly with computation length, meaning fault-tolerant AI architectures must be designed from first principles rather than retrofitted onto existing designs.
\\
\\
Third, the comparison between classical and quantum computational approaches suggests hybrid architectures may be necessary for large-scale intelligence. While classical systems can approach biological efficiency through better engineering, quantum coherence might enable fundamentally different computational paradigms that bypass classical error correction overhead---particularly for pattern recognition and associative memory tasks where approximate solutions suffice.
\\
\\
These physical constraints suggest that the path to artificial general intelligence lies not in scaling current architectures, but in developing new computational paradigms that operate closer to the thermodynamic limits demonstrated by biological systems. The ultimate artificial intelligence may more closely resemble biochemistry than digital logic.

\section{Black hole quantum computation}

\bigskip
Black holes represent the most dense concentrations of matter, energy and information in the universe.  The idea that black holes might function as quantum computers was proposed in [1-2, 14].  A black hole could in principle be programmed to process information in a systematic fashion by fixing the state of the matter that initially forms the hole, and by tailoring the state of infalling matter to program the hole's dynamics. 
As long as black hole evaporation is unitary, then the processed information eventually emerges in the outgoing Hawking radiation. 
\\

We now apply the measures of physical computational complexity presented above to revisit the concept of black hole quantum computation in light of recent progress in understanding black hole dynamics [15-20].   In particular, models of information escape from black holes [15-16] combined with models of fast-scrambling black hole dynamics, e.g., the Hayden-Preskill process [17-18] and the SYK model [19-20], allow a more thorough investigation of the information processing dynamics by which black holes might be able to compute.
\\

Cosmological observations have established that our universe is at or very close to its critical energy density -- the density at which it has just enough energy to expand forever.   Because the critical density for the universe is the same as the density of a black hole, many of our results on computational capacity also apply to the universe as a whole [2].

 \subsection{Computational complexity and black holes}

\bigskip
 A black hole with mass $M$ has Schwarzschild radius $R=2GM/c^2$, entropy $S=4\pi M^2/m_P^2$, and temperature $T = m_P^2 c^2/ 8\pi M$,  where $m_P^2 = \hbar c/G$ is the Planck mass squared, and we have set $k_B = 1$ [1-2, 14-15] [provide a standard reference for black hole evaporation]. 
 Applying the Margolus-Levitin theorem [1], the maximum number of logical operations per second that can be performed by a black hole is $2Mc^2/\pi\hbar$.
 If the energy in the black hole is distributed uniformly among its $4\pi M^2/m_P^2$ quantum bits, then the amount of time it takes each bit to flip (as measured by a distant observer) is $\pi^2 R/c$ [1-2,14]: 
all bits can communicate with each other in the time it takes each one to flip.
The black hole is the ultimate serial computer.  Note that this feature of black hole computation is consistent with the fast scrambling required for the Hayden-Preskill protocol [17-18], and with the all-to-all interactions of the SYK model [19-20].
\\

At first glance, the basic results of physical computational complexity presented above seem to resist the notion that black holes might be efficient information processors. If black hole evaporation is in fact unitary, however, then we can in principal program the black hole by determining the state of the matter and energy that are used to construct the black hole in the first place.   The unitary dynamics of the black hole then processes that information, which is transformed and emitted in the outgoing Hawking radiation. 
\\

The lifetime of a black hole with mass $M$ is
$t_M = 5120 \pi G^2 M^3/\hbar c^4 \propto R^3/\ell_P^2 c$, where $\ell_P^2 = \hbar G/c^3 $ is the Planck length squared.  The lifetime of a solar mass black hole is $t_M \approx 2.140 \times 10^{67}$ years. We see immediately that black holes must be quite small in order to perform useful computation over a human lifespan.   
\\

Having formed and programmed the black hole, we now let it start to evaporate. The initial Hawking radiation escaping from the hole is fully entangled with the hole and is completely random.   That is, each bit emitted increases the entropy of the hole's environment by one bit.   After half the hole's  lifetime, however (the Page time), each bit emitted {\it decreases} the entropy of the environment.   At this point, it is possible in principle to extract the results of the computation that the hole has been programmed to perform by making suitable measurements on the entire Hawking radiation emitted by the hole up to that point [17-18].   In practice, the measurements that need to be made to extract the processed information from the Hawking radiation could be difficult, requiring full knowledge of the internal black hole dynamics [16-18]. Since the black hole lifetime is proportional to $M^3$ and the number of operations per second that it can perform is proportional to $M$, the total number of operations that could be performed by the black hole is equal to $2Mc^2 t_M/\pi\hbar \propto M^4/m_P^4 $.  
\\

The free energy of a black hole is $F = Mc^2 - TS = Mc^2/2$, half the energy of the hole. That is, half of the energy in the hole, $TS=mc^2/2$, is thermal energy that drives the thermal fluctuations in entropic Hawking radiation.  The other half of the hole's energy is free energy and is available for performing computation.

\subsection{Why Programming a Black Hole Might Be Physically Intractable}

\bigskip
Comparing the physics of black hole computation with the measures of physical complexity given in sections 1-3 above, 
we see that errors performed by the black hole during computation are an issue.   In particular, if the computation performed by the hole is generating errors at a constant rate $\epsilon$ per operation, then as noted above at least $-\epsilon \log \epsilon$ bits of entropy are produced per op. The black hole performs $O(M^4/m_P^4)$ ops over its lifetime, but has a maximum information storage capacity/entropy of $O(M^2/m_P^2)$.    The necessity to get rid of the entropy generated by error correction thus requires the error rate per operation to be less than $O(m_P^2/M^2)$. That is, the problem of error correction for even very small black holes appears to be physically intractable, because the physical spatial complexity of the error correction process is greater than the physical spatial complexity of the black hole.
\\
\\

The physical computational complexity of the problem of decoding the information in the Hawking radiation to reveal the results of the black hole's computation may also be very large.    Harlow and Hayden [21] and Aaronson [22] have suggested that the scrambling dynamics of black holes is intrinsically cryptographic, and that the transformation required to unscramble the Hawking radiation has very high computational (and hence physical) complexity.
Indeed, the fast scrambling dynamics and all-to-all couplings of models of black hole dynamics such as the SYK model [19-20] make it hard to imagine -- even in principle -- how one might program a black hole to compute, and to decode the result of the computation from the Hawking radiation.

\section{Conclusion}\

\bigskip
This paper investigated different physical measures of complexity that are analogues of conventional computational measures.   Physical complexity measures the amount of physical resources -- time, energy, information, and free energy -- required to perform physical processes, including computation.   
\\

Since fundamental chemical biological information processing, which takes place at the molecular level, is much more efficient than the information processing in human-built computers, the amount of information processing taking place in living systems vastly outweighs artificial information processing: the biochemical information processing taking place in the living cells of a single human being is the same order of magnitude as the information processing being performed by all electronic computers on earth [23].  
\\

We showed that once the necessity of error correction is included in the accounting of information and entropy, the physical memory space required by a computation, is proportional to the number of logical operations performed in the computation.   We proposed that the physical complexity of a computation is proportional to the amount of free energy required by the computation.  Recasting the computational complexity hierarchy in terms of physical complexity requires us to reexamine the relationships between complexity classes.   
\\

Our analysis suggests that fundamental physical laws impose constraints on what computation is — and perhaps, on what intelligence can become. Whether encoded in the collapse of stars or the dance of biomolecules, the ultimate computer may already be running — not on silicon, but on spacetime itself.

\vskip 1in
\noindent{\bf References:}

\vskip 1cm
\bigskip\noindent[1]
Lloyd S 2000 Ultimate Physical Limits to Computation {\it Nature} {\bf 406} 1047-1054

\bigskip\noindent[2]
Lloyd S 2002 Computational Capacity of the Universe
{\it Physical Review Letters} {\bf 88} 237901

\bigskip\noindent[3] Landauer R 1961 Irreversibility and heat generation in the computing process {\it IBM Journal of Research and Development} {\bf 5} 183–191

\bigskip\noindent[4]
Lecerf Y. 1963 Machines de Turing réversibles {\it Comptes Rendus Hebdomadaires des Séances de L’académie des Sciences} {\bf 257} 2597–2600 (Also available at http://www.cise.ufl.edu/~mpf/rc/Lecerf/lecerf.html)

\bigskip\noindent[5]
Bennett C H 1973 Logical reversibility of computation. {\it IBM Journal of Research and Development} {\bf 17} 525–532

\bigskip\noindent[6]
Bennett C H 1982 The thermodynamics of computation—a review  {\it International Journal of Theoretical Physics} {\bf 21} 905–940

\bigskip\noindent[7]
Fredkin E and Toffoli T 1982 Conservative logic {\it International Journal of Theoretical Physics}  {\bf 21} 219–253

\bigskip\noindent[8]
Benioff P 1980 The computer as a physical system: A microscopic quantum mechanical Hamiltonian model of computers as represented by Turing machines
{\it Journal of Statistical Physics} {\bf 22} 563-591

\bigskip\noindent[9]
Deutsch D 1985 Quantum Theory, the Church-Turing Principle and the Universal Quantum Computer
{\it Proceedings of the Royal Society of London. Series A} {\bf 400} 97-117

\bigskip\noindent[10]
Lloyd S and Pagels H 1988 Thermodynamic depth {\it Annals of Physics} {\bf 188} 186-213

\bigskip\noindent[11]
Turing A 1936 On Computable Numbers, with an Application to the {\it Entscheidungsproblem}
{\it Proceedings of the London Mathematical Society} {\bf 42} 230

\bigskip\noindent[12]
von Neumann J 1958 The Computer and the Brain
Yale University Press
Theory of Self-Reproducing Automata 1966 University of Illinois Press

\bigskip\noindent[13]
Sharma A Czegel D Coachman M Kempes C P, Walker S I and Cronin L 2023
Assembly theory explains and quantifies selection and evolution
{\it Nature}  {\bf 622} 321-328

\bigskip\noindent[14]
Lloyd S and Ng J 2004 Black Hole Computers {\it Scientific American} {\bf 291} 52

\bigskip\noindent[15]
Lloyd S and Preskill J 2014 Unitarity of black hole evaporation in final-state projection models {\it Journal of High Energy Physics} {\bf 2014} 126

\bigskip\noindent[16]
Akers C Engelhardt N and Harlow D 2020 Simple holographic models of black hole evaporation {\it Journal of High Energy
Physics} {\bf 2020} 32

\bigskip\noindent[17]
Hayden P and Preskill J 2007 Black holes as mirrors: quantum information in random subsystems {\it Journal of High Energy Physics} {\bf 09} 120

\bigskip\noindent[18]
Yoshida B and Kitaev A 2017 Efficient decoding for the Hayden-Preskill protocol
arXiv: 1710.03363

\bigskip\noindent[19]
Sachdev S and Ye J 1993 Gapless spin-fluid ground state in a random quantum Heisenberg magnet
{\it Phys. Rev. Lett.} {\bf 70} 3339–42

\bigskip\noindent[20]
Kitaev A 2015 A simple model of quantum holography {\it KITP Strings Seminar and Entanglement
2015 Program} 

\bigskip\noindent[21]
Harlow D and Hayden, P 2013 Quantum computation vs. fire-walls {\it Journal of High Energy Physics} {\bf 6} 1-56

\bigskip\noindent[22]
Aaronson S 2016 The Complexity of Quantum States and Transformations: From Quantum Money to Black Holes arXiv: 1607.05256

\bigskip\noindent[23]
Lloyd S and Reilly M 2025 Natural Intelligence: The Information Processing Power of Life

\end{document}